\documentclass[%
 preprint,
 aps,
 prd,
 floatfix,
]{revtex4-1}

\usepackage{wrapfig}
\usepackage{amssymb}
\usepackage{amsfonts}
\usepackage{amsmath}
\usepackage{lineno}
\usepackage{color}
\usepackage{xcolor}
\usepackage{multirow}
\usepackage{longtable}
\usepackage{bigstrut}
\usepackage{float}

  \usepackage{graphicx}
  \usepackage{epstopdf}
\DeclareGraphicsExtensions{.png,.pdf,.jpg,.mps,.eps}
\graphicspath{{./}}
\def\alr{A_{\rm LR}}
\usepackage{rotate}
\usepackage{latexsym}
\usepackage{ragged2e}
\usepackage{tabularx}
\usepackage{mathtools}
\usepackage{tikz}
\usepackage{slashed}

\def\mev{{\hbox{MeV}}}

\def\alr{A_{\rm LR}}

\newcommand{\bq}{\begin{equation}}
\newcommand{\eq}{\end{equation}}
\newcommand{\bqa}{\begin{eqnarray}}
\newcommand{\eqa}{\end{eqnarray}}

\definecolor{darkblue}{rgb}{0.0, 0.0, 0.55}
\definecolor{darkgreen}{rgb}{0.0, 0.2, 0.13}
\definecolor{darkcandyapplered}{rgb}{0.64, 0.0, 0.0}
\definecolor{color1}{rgb}{0.7,0,0}
\definecolor{color2}{rgb}{0,0,0.7}
\definecolor{color3}{rgb}{0,0,1}
\definecolor{color4}{rgb}{0.8,0,0}
\definecolor{color5}{rgb}{0,0.7,0}

\newcommand{\GeV}{{\text{ GeV}}}

\usepackage{accents}

\newcommand{\vertsp}[1][]{\,#1\vert\, \mathopen{}}

\DeclarePairedDelimiterX\SpAIA[2]{\langle}{\rangle}{{#1}\vertsp[\delimsize]{#2}}
\DeclarePairedDelimiterX\SpAIB[2]{\langle}{]}{{#1}\vertsp[\delimsize]{#2}}
\DeclarePairedDelimiterX\SpBIA[2]{[}{\rangle}{{#1}\vertsp[\delimsize]{#2}}
\DeclarePairedDelimiterX\SpBIB[2]{[}{]}{{#1}\vertsp[\delimsize]{#2}}

\DeclarePairedDelimiterX\SpAIIA[3]{\langle}{\rangle}{{#1}\vertsp[\delimsize]{#2}\vertsp[\delimsize]{#3}}
\DeclarePairedDelimiterX\SpAIIB[3]{\langle}{]}{      {#1}\vertsp[\delimsize]{#2}\vertsp[\delimsize]{#3}}
\DeclarePairedDelimiterX\SpBIIA[3]{[}{\rangle}{      {#1}\vertsp[\delimsize]{#2}\vertsp[\delimsize]{#3}}
\DeclarePairedDelimiterX\SpBIIB[3]{[}{]}{            {#1}\vertsp[\delimsize]{#2}\vertsp[\delimsize]{#3}}

\DeclarePairedDelimiterX\SpIAoAI[2]{{}}{{}}{\SpA[\delimsize]{#1}\aSp[\delimsize]{#2}}
\DeclarePairedDelimiterX\SpIAoBI[2]{{}}{{}}{\SpA[\delimsize]{#1}\bSp[\delimsize]{#2}}
\DeclarePairedDelimiterX\SpIBoAI[2]{{}}{{}}{\SpB[\delimsize]{#1}\aSp[\delimsize]{#2}}
\DeclarePairedDelimiterX\SpIBoBI[2]{{}}{{}}{\SpB[\delimsize]{#1}\bSp[\delimsize]{#2}}

\DeclarePairedDelimiterX\SpAIIIA[4]{\langle}{\rangle}{{#1}\vertsp[\delimsize]{#2}\vertsp[\delimsize]{#3}\vertsp[\delimsize]{#4}}
\DeclarePairedDelimiterX\SpBIIIB[4]{[}{]}{{#1}\vertsp[\delimsize]{#2}\vertsp[\delimsize]{#3}\vertsp[\delimsize]{#4}}

\def\sqs{\sqrt{s}}

\def\sanc{\tt {SANC}}

\def\renesance{\tt {ReneSANCe}}
\def\calchep{\tt {CalcHEP}}
\def\whizard{\tt {WHIZARD}}

\def\sba{$\sigma^{\text{Born}}_{\alpha(0)}$}
\def\sbg{$\sigma^{\text{Born}}_{G_\mu}$}

\def\dbg{$\delta^{\text{Born}}_{G_\mu / \alpha(0)}$}

\def\swa{$\sigma^{\text{weak}}_{\alpha(0)}$}
\def\swg{$\sigma^{\text{weak}}_{G_\mu}$}

\def\dwg{$\delta^{\text{weak}}_{G_\mu / \alpha(0)}$}

\begin{document}

\title{One-loop electroweak radiative corrections to polarized $e^+e^- \to Z Z$ process}

\author{S.~Bondarenko}
\altaffiliation[Also at ]{
  Dubna State University, Dubna 141980, Russia}
\email{bondarenko@jinr.ru}
\affiliation{
  Bogoliubov Laboratory of Theoretical Physics, Joint Institute for Nuclear Research,
  Dubna, 141980 Russia}

\author{Ya.\,Dydyshka}
\altaffiliation[Also at ]{
  Institute for Nuclear Problems,
  Belarusian State University, Minsk, 220006  Belarus}

\author{L.~Kalinovskaya}

\author{R. Sadykov}

\author{V.~Yermolchyk}
\altaffiliation[Also at ]{
  Institute for Nuclear Problems,
  Belarusian State University, Minsk, 220006  Belarus}
\affiliation{%
  Dzhelepov Laboratory of Nuclear Problems, Joint Institute for Nuclear Research, Dubna,
  141980 Russia
}%


\begin{abstract}
In the paper we recalculate and discuss high-precision theoretical predictions for cross sections of the process $e^+e^- \to  Z Z$.
We assume
a complete one-loop implementation and a possibility of estimating
the initial state polarizations, as well as the full-phase calculation.
Numerical results are provided by our Monte-Carlo tools {\tt MCSANC} integrator and {\tt ReneSANCe} generator for typical energies and degrees of polarization of ILC and CLIC projects
in two $\alpha(0)$ and G$_{\mu}$ electroweak schemes.
\end{abstract}

\maketitle

\section{Introduction}
The process of $Z$ pair production together
with the process $e^+e^- \to \gamma Z$ is the main background for the reaction $e^+e^- \to ZH$ at the center-of-mass system (c.m.) energy of
250 GeV in the Higgs boson measurement method.
This method should identify Higgs boson events independent of the decay mode
allowing the measurement of the total cross section for Higgs production.
Recently we have estimated theoretical uncertainties for
the polarized annihilation $e^+e^- \to ZH$ \cite{Bondarenko:2018sgg,Arbuzov:2021zjc}
and $e^+e^- \to \gamma Z$ \cite{Bondarenko:2024jetplaz} in the same way, i.e., by describing of the one-loop level using the massive helicity approach in the full phase space.

To our knowledge, the QED and electroweak (EW) corrections to the unpolarized $Z$ boson pair production have previously been
calculated  only in ~\cite{Denner:1988tv,Gounaris:2002za,Demirci:2022lmr}.

In this article we revise the uncertainties in the theoretical
interpretation of the process
\bqa
e^{+}(p_1,\chi_1) + e^{-}(p_2,\chi_2)
\rightarrow  Z(p_3,\chi_3) + Z(p_4,\chi_4) \  (+ \gamma(p_5,\chi_5)).
\label{Reac_eeZZ}
\eqa
For the virtual part, we discussed analytic expressions for the covariant amplitude, tensor structures,
helicity amplitudes
and presented them in a compact form in~\cite{Bardin:2005dp}.
The contribution of the hard real photon emission is
obtained by direct squaring of the matrix element.

In this paper we extend the research 
and evaluate the complete one-loop corrections supplemented
by higher-order QED contributions in the leading logarithmic 
approximation (LLA)
by structure function approach~\cite{Kuraev:1985hb}.
The impact of the initial-state radiation (photons and
pairs) is analyzed order by order. 
We used known expressions for contributions of the collinear electron
structure function of the orders 
${\cal O}(\alpha^nL^n)_{\gamma, pairs}, n=2-4$
 for photons and pairs~\cite{Arbuzov:2021zjc}. 
Based on this background we analyse
 the size of the radiative corrections and different higher-order contributions.
 Presumably, 
 to evaluate ISR  QED corrections
 the exponentiated procedure is more suitable for Monte Carlo simulations, while the
order-by-order one can be used for benchmarks and cross-checks.
We intend to use 
a parton shower for QED based on quantum density matrix factorization that was  proposed by Nagy-Soper~\cite{Nagy:2008eq,Nagy:2014mqa}. This approach allows resummation of all collinear logarithms taking into account spin correlations and can be a valuable alternative for older YFS-based algorithms~\cite{Yennie:1961ad,Jadach:1998jb}.

This work explores the influence of the initial
beam polarization at the planned experiments on ILC, FCC, CEPC
for the reaction $ e^+e^- \to ZZ$ with both $Z$ bosons on-mass-shell.
Since this issue has not been studied before, 
we decided to investigate it carefully and to publish the results missing in literature.

We consider a narrow-width cascade using Born and one-loop $e^+e^- \to ZZ$
and $Z \to  \mu^+\mu^-$ results as
a rough estimation of the partial $e^+e^- \to ZZ 
\stackrel{\text{n.w.}}{\to} 2\mu^+ 2\mu^-$ channel. 
The gauge-invariant analytical results of initial state QED corrections to off-shell vector boson pair production were given in \cite{Bardin:1996jw,Jadach:1996hh}. 
The corrections in which
the two $W$- or $Z$-boson resonances are not independent due to the exchange of
soft photons between the different subprocesses (non-factorizable corrections) are estimated in \cite{Denner:1998rh}.

Whenever possible, we compare our results with those described in literature.
Polarized tree-level cross sections
(the Born and hard photon bremsstrahlung) are compared with the {\calchep}~\cite{Belyaev:2012qa}
and {\whizard}~\cite{Kilian:2007gr,Kilian:2018onl} results;
the weak and QED parts, with \cite{Denner:1988tv};
the NLO level, with the results \cite{Demirci:2022lmr}.

Numerical results are presented for the total and differential cross sections 
that are functions of the cosine of the scattering angle, 
and for the relative corrections in the $\alpha(0)$ and $G_\mu$ EW schemes with an estimation of the polarization effects of the initial states.

All calculations were carried out using the {\tt MCSANC} integrator
and the {\renesance} generator~\cite{SADYKOV2020107445}
which allow to evaluate the arbitrary differential cross sections and separate particular contributions.

The article is organized as follows. 
Section~\ref{PAP} describes the stage of the calculation of the polarized cross sections at the complete one-loop EW level. We consistently set out the relevant components of the one-loop cross section within the helicity approach.
In Section~\ref{NumResultsComp}, 
the tuned comparison with third-party codes is presented for tree and one-loop levels.
The corresponding numerical results
are given for the total, differential cross sections,
relative corrections with an estimation of polarization effects,
left-right asymmetry and narrow-width approximation for decay channel.

In Section~\ref{sect_Concl} the discussion and conclusions are given.

\section{Differential cross section \label{PAP}}

The cross section of any reaction  $e^+e^-$ annihilation
with longitudinal  electron beam polarizations $P_{e^-}$ and positron beam polarization $P_{e^+}$ is computed from the four possible pure 
helicity cross sections.
To study the case of the longitudinal polarization with degrees $P_{e^+}$ and $P_{e^-}$ 
we  make a formal application of Eq.~(1.15) from
~\cite{MoortgatPick:2005cw}:
\begin{equation}
\sigma{(P_{e^+},P_{e^-})} = 
\frac{1}{4}\sum_{\chi_1,\chi_2}(1+\chi_1 P_{e^+})
(1+\chi_2 P_{e^-}) \sigma_{\chi_1\chi_2},
\label{eq1}
\end{equation}
where $\chi_{i} = -1(+1)$ corresponds to the particle $i$ with the left (right) helicity.

The cross section of the process at the one-loop level can be divided into four parts:
\begin{eqnarray}
  \sigma^{\text{one-loop}}_{{ \chi_1}{ \chi_2}{}{}{}} =  \sigma^{\mathrm{Born}}_{{ \chi_1}{ \chi_2}{}{}{}}
        + \sigma^{\mathrm{virt}}_{{ \chi_1}{ \chi_2}{}{}{}}(\lambda)
        + \sigma^{\mathrm{soft}}_{{ \chi_1}{ \chi_2}{}{}{}}({\lambda},{\bar\omega})
        + \sigma^{\mathrm{hard}}_{{ \chi_1}{ \chi_2}{}{}{}}({\bar\omega}).
\end{eqnarray}
Here $\sigma^{\mathrm{Born}}$ is the Born cross section,
$\sigma^{\mathrm{virt}}=\sigma^{\mathrm{QED}}+\sigma^{\mathrm{weak}}$ is the contribution of virtual (loop) corrections,
$\sigma^{\mathrm{soft(hard)}}$ is the soft (hard) photon emission contribution
(the hard photon energy { $E_{\gamma} > {\bar\omega} = \omega \sqs/2$}).
We divide the virtual part into two gauge invariant subsets: $\sigma^{\mathrm{QED}}$ and $\sigma^{\mathrm{weak}}$. To the QED contribution we refer all diagrams in which there is an exchange of at least one photon. The rest is the weak part.
 Auxiliary parameters {$\lambda$} ("photon mass") and
 {${\bar\omega}$} (soft-hard separator) are canceled after summation.
The cancellation is controlled numerically by calculating the cross section at several values of the $\lambda$ and $\bar\omega$ parameters.
When calculating the emission of real photons we keep the electron masses to regularize the collinear divergences.

\section{Numerical results and comparisons
\label{NumResultsComp}}

For numerical evaluations we used the following  setup of input parameters:
\begin{eqnarray}
\alpha^{-1}(0) &=& 137.035999084,
\\
M_W &=& 80.379 \; \GeV, \quad
M_Z = 91.1876 \; \GeV, \quad M_H = 125 \; \GeV,
\nonumber\\
\Gamma_Z &=& 2.4952 \; \GeV, \quad m_e = 0.51099895 \; \mev,
\nonumber\\
m_\mu &=& 0.1056583745 \; \GeV, \quad m_\tau = 1.77686 \; \GeV,
\nonumber\\
m_d &=& 0.083 \; \GeV, \quad m_s = 0.215 \; \GeV,
\nonumber\\
m_b &=& 4.7 \; \GeV, \quad m_u = 0.062 \; \GeV,
\nonumber\\
m_c &=& 1.5 \; \GeV, \quad m_t = 172.76 \; \GeV,
\nonumber
\end{eqnarray}
and set of c.m. energies
\begin{eqnarray}
\label{set_energy}
\sqs= 250, 500 ~\mbox{and} ~1000 ~\mbox{GeV}.
\end{eqnarray}
The following longitudinally polarized states are considered:
\begin{eqnarray}
&&(P_{e^+}, P_{e^-}) = (0, 0),(-1, +1),(+1, -1),\\
&&(-0.3, +0.8), (+0.3, -0.8), (0, +0.8), (0, -0.8). \nonumber
\end{eqnarray}
where ($P_{e^+}, P_{e^-}$) are the positron and electron beam polarizations.

Original calculations were performed without angular cuts, while for comparison
we used the cuts from considered papers.

\subsection{Comparison with other codes}

\subsubsection{The tree level}

The agreement in 5 digits was found for the results for the
Born cross section
with the codes {\tt CalcHEP} and {\tt WHIZARD}, so we omitted the corresponding table.

The results of the comparison for the hard photon bremsstrahlung
with the only cut on the
photon energy  $E_\gamma > \bar{\omega} = 10^{-4} \sqrt{s}/2$
within the $\alpha(0)$ EW scheme for 
 unpolarized and fully polarized initial beams
are given in Table~\ref{Table:tuned_hard_SCW}.
The agreement within four digits is demonstrated.

\begin{table}[!ht]
\begin{center}
\caption{\label{Table:hard_350500}
The tuned triple comparison
between the {\sanc} (first line), {\calchep} (second line) and {\whizard} (third line) hard photon bremsstrahlung
contributions $\sigma^{\text{hard}}$ (fb) to polarized $e^+e^- \to Z Z(\gamma)$ scattering
for various degrees of polarization and energies
  }
\label{Table:tuned_hard_SCW}
\begin{tabular}{lccc}
\hline\hline
$P_{e^+}$, $P_{e^-}$ & $0, 0$ & $-1, +1$ & $+1, -1$ \\
\hline 
\multicolumn{4}{c}{$\sqs = 250$ GeV}       \\
S & 985.3(1) & 1165.8(1) & 2774.2(1)  \\
C & 985.2(1) & 1165.7(1) & 2774.1(1)  \\
W & 985.4(1) & 1166.0(1) & 2774.3(1)  \\
\multicolumn{4}{c}{$\sqs = 500$ GeV}       \\
S & 433.1(1) & 511.7(1) & 1217.8(1) \\
C & 433.1(1) & 511.9(1) & 1217.6(1) \\
W & 433.2(1) & 511.7(1) & 1218.1(1) \\
\multicolumn{4}{c}{$\sqs = 1000$ GeV}      \\
S & 173.0(1) & 204.2(1) & 486.0(1) \\
C & 173.0(1) & 204.3(1) & 486.1(1) \\
W & 173.0(1) & 204.2(1) & 486.0(1)  \\
\hline\hline
\end{tabular}
\end{center}
\end{table}

\subsubsection{The one-loop level} 

\underline{Comparison with~\cite{Denner:1988tv}.}

\begin{figure}[h!]
    \includegraphics[width=0.5\textwidth]{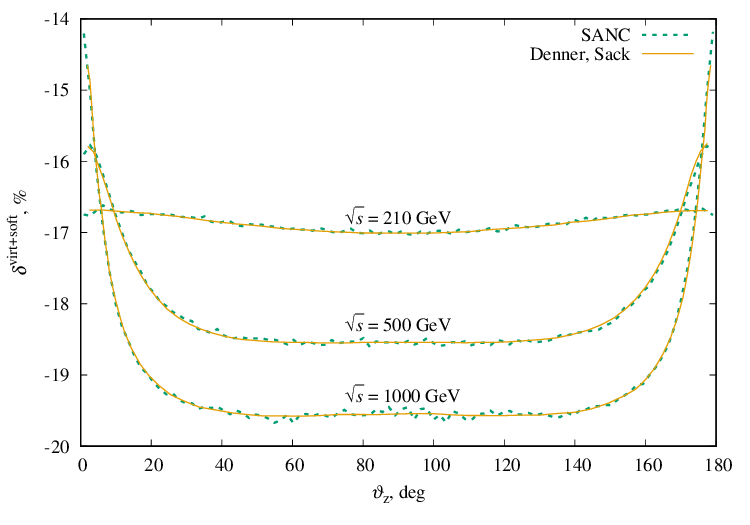}
    \caption{\label{fig:comp-delta-qed}
    Tuned comparison between the {\sanc} and
    \cite{Denner:1988tv} results
    for the relative QED corrections $\delta^{\rm virt + soft}({\omega}=0.1)$
    }
\end{figure}

\begin{figure}[h!]
    \includegraphics[width=0.45\textwidth]{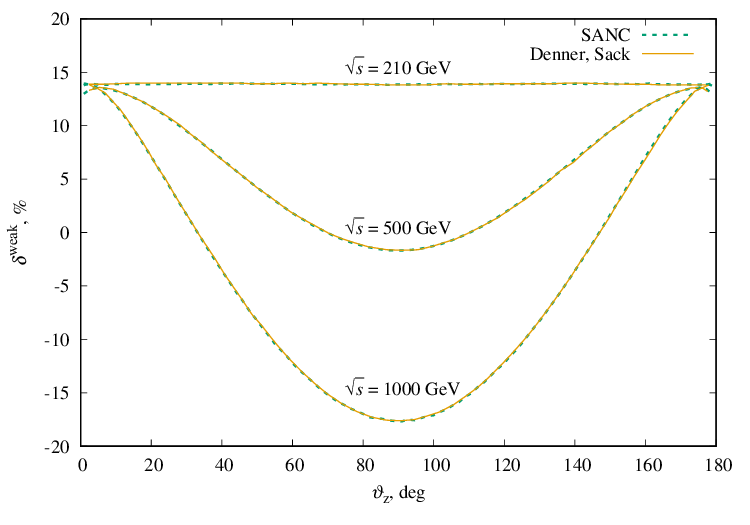}
    \caption{\label{fig:comp-delta-weak}
    Tuned comparison between the {\sanc} and \cite{Denner:1988tv} results
    for the relative weak corrections $\delta^{\rm weak}$
    }
\end{figure}

\begin{figure}[h!]
    \includegraphics[width=0.5\textwidth]{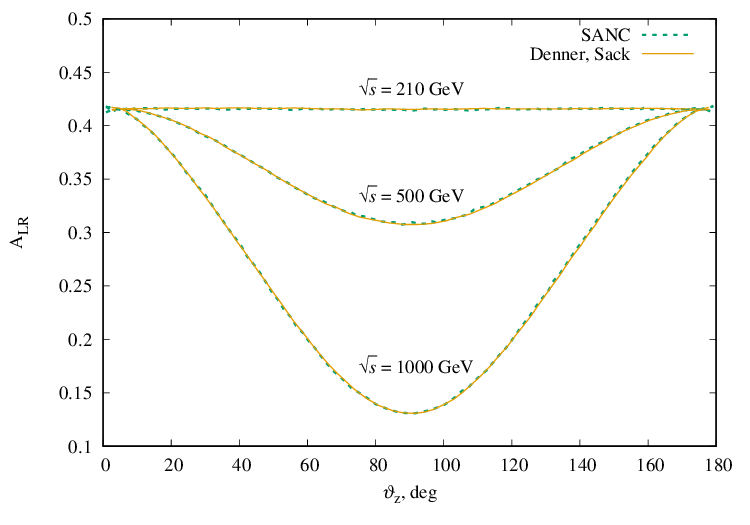}
   \caption{\label{fig:comp-alr-nlo}
   Tuned comparison between the {\sanc} and \cite{Denner:1988tv} results
   for $A_{\rm LR}$
   }
\end{figure}

We also made a separate comparison between {\tt SANC} and \cite{Denner:1988tv} for the QED virtual part with the soft photon contribution $\delta^{\rm virt+soft}$ and the virtual weak contribution to the NLO calculations $\delta^{weak}$.
In our previous paper~\cite{Bardin:2007zz}  
we found an excellent agreement with the virtual weak contribution numbers from the fourth column in Table 1 ~\cite{Denner:1988tv} with original setups and cuts.
In this research we additionally compare the corresponding
angular distributions of the QED virtual part with soft photon relative corrections $\delta^{\rm virt+soft}$
(Fig.~\ref{fig:comp-delta-qed}) and
the virtual weak relative corrections $\delta^{\rm weak}$ (Fig.~\ref{fig:comp-delta-weak}).
The calculated $\alr$ asymmetry of the QED and weak virtual parts with the soft photon contribution is presented in Fig.~\ref{fig:comp-alr-nlo}.
The results show a very good agreement with those given in~Figs.~9, 11 and 12 of~\cite{Denner:1988tv}.

\underline{Comparison with~\cite{Demirci:2022lmr}.}

In~\cite{Demirci:2022lmr} the calculations of complete one-loop results of 
the process ~(\ref{Reac_eeZZ}) are given
with allowance for the longitudinal polarization of the initial beams.
We compare our results with those obtained in~\cite{Demirci:2022lmr} with the original cuts and input parameters for the unpolarized case.

At the tree level (Born and hard-photon bremsstrahlung cross sections) 
we agree with the results in Table~I ~\cite{Demirci:2022lmr} within 4-5 digits. 

For the NLO results, we found a difference in the total relative corrections 
at the c.m. energies (\ref{set_energy}),
namely,  $-1.20(1)$\%, +6.77(1)\%, +6.54(1)\%, respectively (should be compared with Table II of~\cite{Demirci:2022lmr}).

Since the definitions of QED and weak subsets of the one-loop diagrams
differ in the {\tt SANC} system (see, for example,~\cite{Bardin:1998nm} and~\cite{Denner:2019vbn}) and those given in~\cite{Demirci:2022lmr}, 
it is impossible to compare the separate QED and weak contributions.
We also see a difference in the angular distributions of the unpolarized cross sections.

It should also be noted that in~\cite{Demirci:2022lmr} the inconsistent definition of the G$_\mu$ EW scheme is used. To avoid double counting in the G$_\mu$ scheme, the subtraction of the parameter $\Delta r$ in a one-loop precision should be done along with changing the fine structure constant $\alpha(0) \to \alpha_{{\rm G}_\mu}$ (see, for instance,~\cite{Bredenstein:2006rh,Denner:2019vbn}). 
This gives about $-(5-6)$\% to the relative corrections of the virtual contribution in the G$_\mu$ EW scheme. 

\subsection{Total cross sections}

Corresponding unpolarized/polarized results for the Born and complete one-loop EW cross sections
(in picobarns), as well as for relative corrections,
are presented in Table~\ref{Table:total-unpol-pol}.
Relative corrections $\delta^i$ are computed as the ratios (in percent)
of the corresponding
RC contributions to the Born level cross section for three energies.  
We show  only the components $\sigma_{-+}$ and $\sigma_{+-}$ because even in cases of a partly polarized initial state, the polarized cross sections are mainly determined by these components.

\begin{center}
\begin{table}[ht]
\caption{\label{Table:total-unpol-pol}
Integrated Born and one-loop cross sections and
relative corrections 
for unpolarized and polarized initial beams 
at the c.m. energies
(\ref{set_energy}).
}
\begin{tabular}{lccccccc}
\hline\hline
$P_{e^+}, P_{e^-}$           & $0,0$         &$-1,+1$    & $+1,-1$  & $+0.3,-0.8$ & $-0.3,+0.8$ & $0,-0.8$  & $0,+0.8$ \\
\hline 
\multicolumn{8}{c}{$\sqs = 250$~GeV}\\
$\sigma^{\rm Born}$, pb      & 1.0198(1)  & 1.2070(1) & 2.8722(1) & 1.7225(1) & 0.80661(1) & 1.3529(1) &0.68675(1)\\
$\sigma^{\rm NLO}$, pb       & 1.0087(1)  & 1.4717(1) & 2.5625(1) & 1.5508(1) & 0.95079(3) & 1.2270(1) &0.79067(3)\\
$\delta^{\rm NLO}$, \%       & $-1.08(1)$ & 21.93(1)  &$-10.78(2)$&$-9.97(1)$ & 17.88(1)   & $-9.30(1)$&15.14(1)\\
$\delta^{\rm QED}$, \%       & $-1.36(1)$ & $-1.39(1)$&$-1.39(1)$ &$-1.38(1)$ &$-1.37(1)$  &$-1.37(1)$ &$-1.34(1)$\\
$\delta^{\rm weak}$, \%      & 0.29(2)    & 23.32(1)   & $-9.39(1)$ & $-8.59(1)$ & 19.24(2) & $-7.93(1)$& 16.48(2) \\
\multicolumn{8}{c}{$\sqs = 500$~GeV}\\
$\sigma^{\rm Born}$, pb      & 0.38530(1) & 0.45604(1)& 1.08518(2) & 0.65079(1) & 0.30476(1)& 0.51115(1) &0.25947(1)\\
$\sigma^{\rm NLO}$, pb       & 0.41320(2) & 0.69420(2)& 1.04592(3) & 0.63358(3) & 0.39058(2)& 0.50159(2) &0.32486(2)\\
$\delta^{\rm NLO}$, \%       & 7.24(1)    & 32.49(1)  & $-3.62(1)$ & $-2.65(1)$ & 28.16(1)  & $-1.87(1)$ &25.20(1)\\
$\delta^{\rm QED}$, \%       & 9.49(1)    & 9.32(1)   & 9.32(1)    & 9.40(1)    & 9.49(1)   & 9.46(1)    & 9.58(1)\\
$\delta^{\rm weak}$, \%      & $-2.25(1)$ & 23.17(1)  & $-12.94(1)$& $-12.05(1)$ & 18.67(1) & -11.33(1)  & 15.62(1)\\
\multicolumn{8}{c}{$\sqs = 1000$~GeV}\\
$\sigma^{\rm Born}$, pb      & 0.14044(1) & 0.16622(1)& 0.39556(2) &0.23722(1)   &0.11108(1)& 0.18631(1) &0.94571(1)\\
$\sigma^{\rm NLO}$, pb       & 0.15614(2) & 0.23058(1)& 0.39217(2) & 0.23784(2)  &0.14896(1)& 0.18846(2) &0.12382(1)\\
$\delta^{\rm NLO}$, \%       & 11.17(1)   & 38.72(1)  & -0.85(1)   & 0.27(1)     & 34.09(2) &1.16(1)     & 30.92(2)\\
$\delta^{\rm QED}$, \%       & 15.90(1)   & 15.58(1)  & 15.59(1)   & 15.73(1)    & 15.89(1) & 15.82(1)   & 16.06(1)\\
$\delta^{\rm weak}$, \%      & $-4.72(1)$ & 23.13(1)  & $-16.43(1)$& $-15.46(1)$ &18.21(2)  &$-14.67(1)$ & 14.87(2)\\
\hline\hline
\end{tabular}
\end{table}
\end{center}

It is seen that for 
the c.m. energies (\ref{set_energy}),
the weak relative corrections for mostly positive electron polarization are positive and have very low energy dependence. For ($P_{e^+}, P_{e^-}) = (-1,+1$) they are practically constant. 

In the case of mostly negative electron polarization, the weak relative corrections are negative and highly energy-dependent.

The QED relative corrections strongly depend on the energy and very weakly on the degree of initial beam polarizations.

To estimate theoretical uncertainty, we carry out calculations in two EW schemes.
The integrated cross sections for the weak corrections in the $\alpha(0)$ and $G_\mu$ schemes and their relative difference
\begin{eqnarray}
\delta_{G_\mu/\alpha(0)} = \frac{\sigma_{G_\mu}}{\sigma_{\alpha(0)}} - 1,\, \%
\label{rgmual0}
\end{eqnarray}
are presented in Table~\ref{Table:delta_350ewscheme}. 

As is well known, the difference between two EW schemes in the LO is just the
ratio of the EW couplings and gives $\delta^{\rm LO}_{G_\mu/\alpha(0)} = 7.5\%$. 
As seen in the table, the weak contribution reduces the difference
to about
1\% at the energy of 250 GeV, 0.7\% at 500 GeV and 0.4\% at 1000 GeV. 
These ratios~(\ref{rgmual0}) 
show stabilization of the results and can be considered as
an estimation of the theoretical uncertainty of weak contributions, 
which is in line with additional corrections of two and more loops.

\begin{table}[ht]
\caption{
\label{Table:delta_350ewscheme}Integrated Born and weak contributions 
to the cross section corrections in two EW schemes,
$\alpha(0)$ and $G_\mu$,
at the c.m. energies
(\ref{set_energy})
}
\begin{center}
\begin{tabular}{lccc}
\hline\hline
$\sqs$, GeV & 250         & 500       & 1000\\
\hline
{\sba, pb}      & 1.0198(1)  & 0.38530(1) & 0.14044(1) \\
{\sbg, pb}      & 1.0961(1)  & 0.41412(1) & 0.15095(1)\\
{\dbg, \%}      & 7.48(1)    & 7.48(1)    & 7.48(1)\\
{\swa, pb}      & 1.0227(1)  & 0.37663(1) & 0.13381(1)\\
{\swg, pb}      & 1.0323(1)  & 0.37914(1) & 0.13433(1) \\
{\dwg, \%}      & 0.94(1)    & 0.66(1)    & 0.39(1)\\
\hline\hline
\end{tabular}
\end{center}
\end{table}

\subsection{Multiple photon ISR relative corrections}

We evaluate ISR corrections
to high-energy processes
in the channel electron-positron annihilation
within
the LLA using the QED structure function formalism
\cite{Kuraev:1985hb}. For corrections of this kind
the large logarithm corresponds to $L = \ln(s/m^2_e)$, where the total c.m.
energy $\sqrt{s}$ is chosen as the factorization scale.
In Table~\ref{Table:LLAQED5} we show the ISR corrections of different order of ${\cal O}(\alpha^nL^n), n=2-3$ in the leading logarithmic approximation
for the c.m. energies (\ref{set_energy}) in the $\alpha(0)$ EW scheme.
\begin{table}[ht]
\caption{\label{Table:LLAQED5}
Multiple photon ISR relative corrections $\delta$ ($\%$) in the LLA approximation
the set c.m. energies (\ref{set_energy})
}
\centering
\begin{tabular}{lccc}
\hline\hline
$\sqs$, GeV                          & 250  & 500 & 1000\\
\hline
$\mathcal{O}(\alpha L)   $, $\gamma$     & $-2.436(1)$   & $+8.074(1)$  & $+13.938(1)$\\
$\mathcal{O}(\alpha^2L^2)$, $\gamma$     & $-0.692(1)$   & $-0.268(1)$  & $+0.229(1)$\\
$\mathcal{O}(\alpha^2L^2)$, $e^+e^-$     & $-0.013(1)$   & $+0.324(1)$  & $+1.516(1)$\\
$\mathcal{O}(\alpha^2L^2)$, $\mu^+\mu^-$ & $-0.008(1)$   & $+0.199(1)$  & $+0.958(1)$\\
$\mathcal{O}(\alpha^3L^3)$, $\gamma$     & $+0.034(1)$   & $-0.014(1)$  & $-0.016(1)$\\
$\mathcal{O}(\alpha^3L^3)$, $e^+e^-$     & $-0.017(1)$   & $-0.022(1)$  & $-0.051(1)$\\
$\mathcal{O}(\alpha^3L^3)$, $\mu^+\mu^-$ & $-0.010(1)$   & $-0.013(1)$  & $-0.033(1)$\\
$\mathcal{O}(\alpha^4L^4)$, $\gamma$     & $<0.001$   & $<0.001$  & $<0.001$\\
\hline\hline
\end{tabular}
\end{table}
To illustrate the trends of the ISR contribution behaviour, we present separate distributions for each ${\cal O}(\alpha^nL^n)$ term.
When considering corrections in LLA, we see that it is certainly sufficient to take into account corrections up to the third order.
It is seen that the corrections for
the sum of all considered orders of the ISR terms $\sum_{n=2}^3{\cal O}(\alpha^nL^n)$
are about -0.706\% for the c.m. energy $\sqs=250$~GeV and about +0.206\% (+2.603\%) for the c.m. energy $\sqs=500~(1000)$~GeV. 
For the c.m. energy $\sqs=250$~GeV the most significant contribution in LLA 
is of course the photonic one of the order ${\cal O}(\alpha^2 L^2)$. 
For the c.m. energy $\sqs=500~(1000)$~GeV the dominant contributions of the second order are about $+0.324\% ~(1.516\%)$ for $e^+e^-$-pairs.

\subsection{Differential distributions}

\subsubsection{Angular dependence}

\begin{figure}[h!]
\begin{center}
\includegraphics[width=0.5\linewidth]{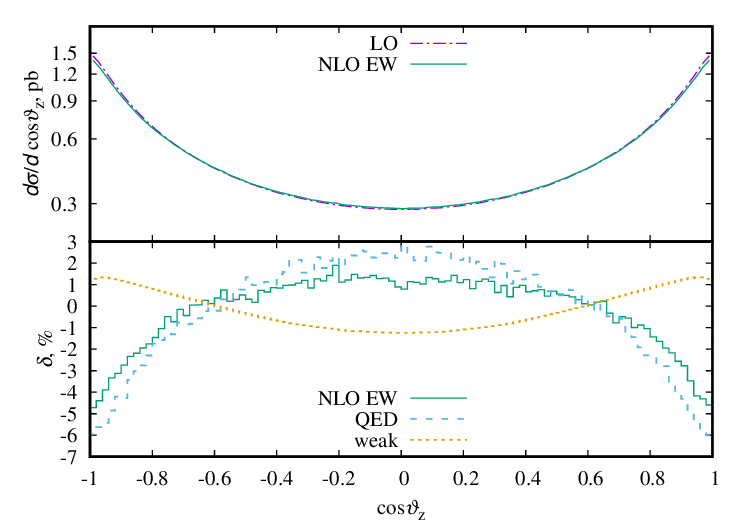}
\caption{
{LO and EW NLO (in parts) cross sections and relative corrections
at $\sqs = 250$~GeV
for unpolarized initial beams}
}
\label{fig:250-cos3}
\end{center}
\end{figure}

\begin{figure}[h!]
\begin{center}
\includegraphics[width=0.5\linewidth]{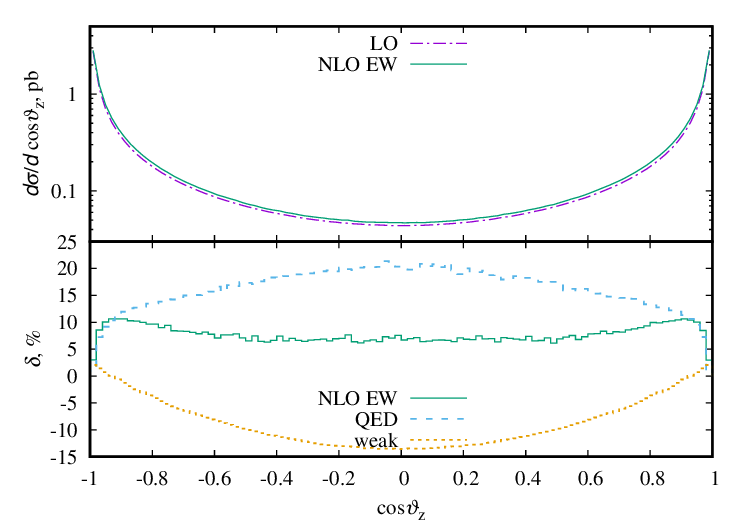}
\caption{The same as in Fig.~\ref{fig:250-cos3} but for $\sqs = 500$~GeV}
\label{fig:500-cos3}
\end{center}
\end{figure}

\begin{figure}[!h]
\begin{center}
\includegraphics[width=0.5\linewidth]{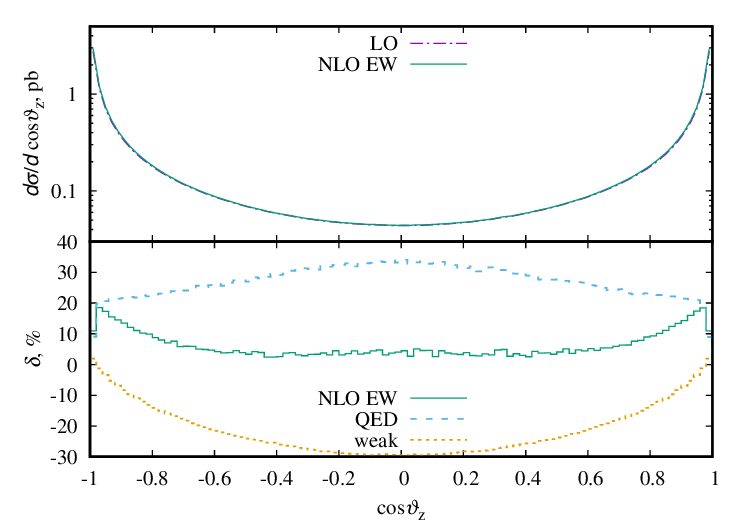}
\caption{The same as in Fig.~\ref{fig:250-cos3} but for $\sqs = 1000$~GeV}
\label{fig:1000-cos3}
\end{center}
\end{figure}

Figures~\ref{fig:250-cos3}, \ref{fig:500-cos3} and \ref{fig:1000-cos3}
show the angular dependence
of the unpolarized cross sections [Born and one-loop level in the $\alpha(0)$ EW scheme] as well as QED and weak relative corrections.
The $\vartheta_Z$ is the angle between the initial
positron $e^+(p_1)$ and any $Z$-boson.

For all c.m. energies
the minimum of the Born and one-loop cross
sections is at zero
(the dependence is symmetric about zero)
while the maximum is in the corners $\cos \vartheta_Z = \pm 1$. 

At $\sqs = 250$ GeV the QED relative corrections dominate and only slightly
change by weak corrections. At $\sqs = 500$ and 1000 GeV both corrections
are large and a strong compensation occurs.

\subsubsection{Energy dependence}
 
\begin{figure}[h!]
\begin{center}
\includegraphics[width=0.5\textwidth]{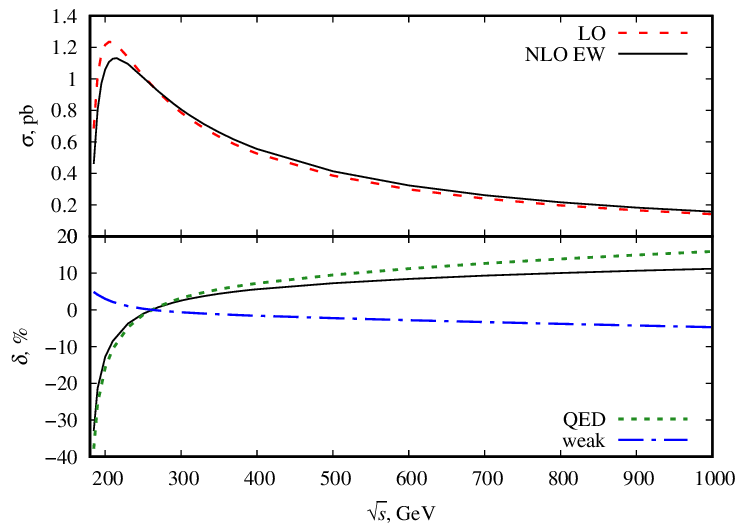}
\end{center}
\caption{
\label{fig:eeZZdelta}
The LO and NLO EW corrected unpolarized cross sections 
and the relative corrections in parts 
as a function of the c.m. energy.
}
\end{figure}

The LO and NLO EW corrected unpolarized cross sections
and the relative corrections in the parts (QED and weak)
as a function of the c.m. energy are shown in Fig.~\ref{fig:eeZZdelta}.

In the c.m. energy range from the threshold to 1000 GeV, the QED corrections
dominate except for the point near $\sqs = 260-270$ GeV, where the QED and weak corrections
are equal to each other. Below this point, the total relative corrections are negative,
then become positive and reach about 10\% at $\sqs = 1000$ GeV.

\subsection{Asymmetries}

All one-loop corrections for this process are symmetric in the
$u \leftrightarrow t$ exchange, so that there is no forward-backward asymmetry which could be easier observed without polarization.

The left-right asymmetry $\alr$ for the fully polarized case is defined as follows:
\bqa
\alr = 
 \frac{\sigma_{\mathrm {LR}}-\sigma_{\mathrm {RL}}}       
 {\sigma_{\mathrm {LR}}+\sigma_{\mathrm {LL}}+\sigma_{\mathrm {RL}}+\sigma_{\mathrm {RR}}},
 \label{alr1}
\eqa
since to cross section $\sigma_{\mathrm {LL,RR}}$ only hard real photons account,
and $\sigma_{\mathrm {LR}}$ 
and $\sigma_{\mathrm {RL}}$ being the cross sections for the fully polarized
electron-positron $e^-_{\mathrm L}e^+_{\mathrm R}$ 
and $e^-_{\mathrm R}e^+_{\mathrm L}$ initial states, respectively.

In the case of partially polarized initial beams, the asymmetry can be written as
\bqa
\alr = 
 \frac{\sigma(P_{e^+},P_{e^-})-\sigma(-P_{e^+},-P_{e^-})}
      {\sigma(P_{e^+},P_{e^-})+\sigma(-P_{e^+},-P_{e^-})}.
 \label{alr2}
\eqa
At the Born level $\alr$ is constant \cite{Denner:1988tv}.

\begin{figure}[!h]
\begin{center}
\includegraphics[width=0.5\linewidth]{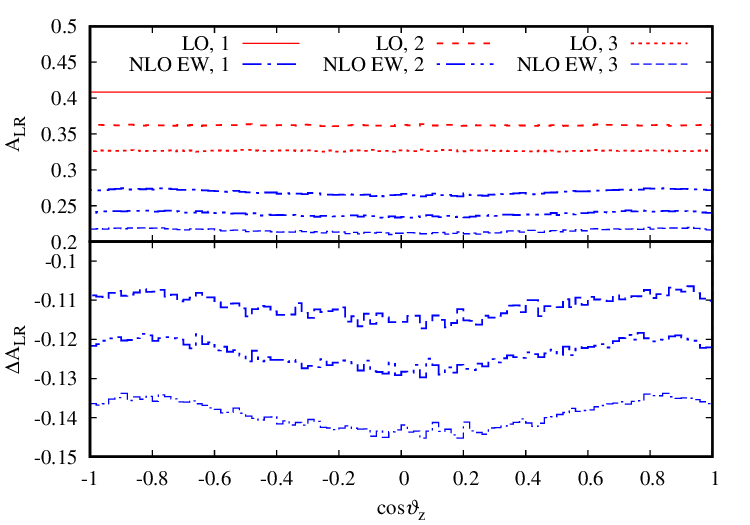}
\caption{
The asymmetry $\alr$  in the Born and one-loop approximations 
at c.m. energy $\sqs = 250$~GeV
for fully polarized and partially polarized initial beams
vs. the cosine of the scattering angle.
}
\label{fig_alr:250}
\end{center}
\end{figure}

\begin{figure}[!h]
\begin{center}
\includegraphics[width=0.5\linewidth]{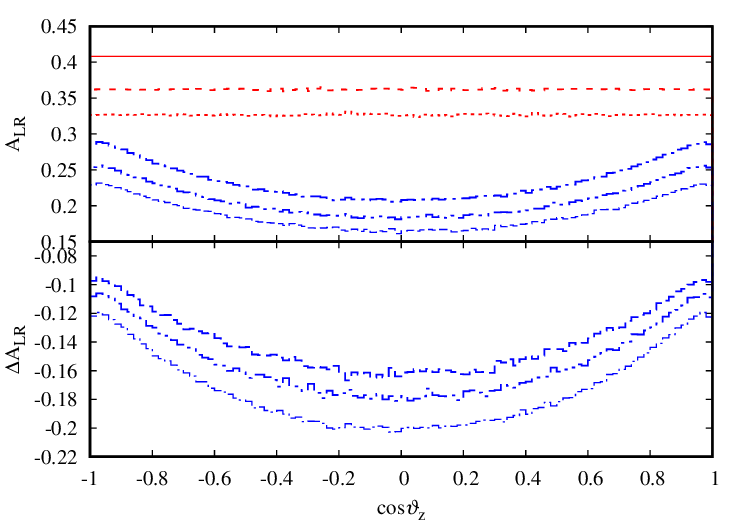}
\caption{
The same as in Fig.~\ref{fig_alr:250}
but for c.m. energy $\sqs = 500$~GeV.
}
\label{fig_alr:500}
\end{center}
\end{figure}

\begin{figure}[!h]
\begin{center}
\includegraphics[width=0.5\linewidth]{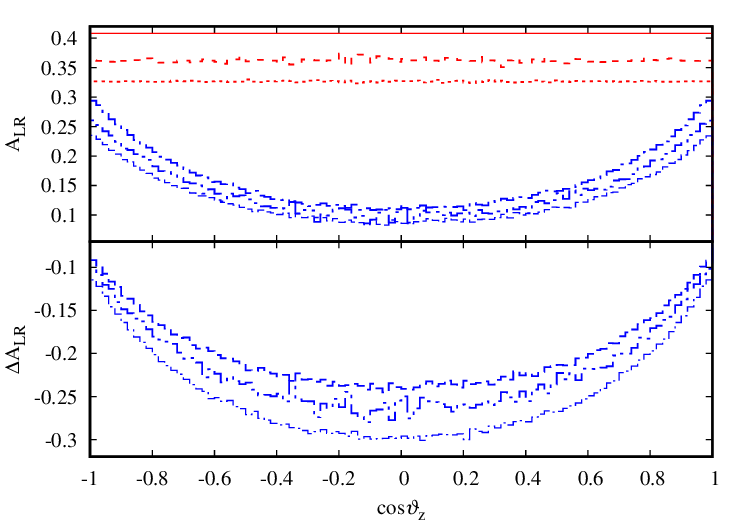}
\caption{
The same as in Fig.~\ref{fig_alr:250}
but for c.m. energy $\sqs = 1000$~GeV.
}
\label{fig_alr:1000}
\end{center}
\end{figure}

In Figs.~\ref{fig_alr:250}-\ref{fig_alr:1000}
the left-right asymmetry distributions for Born and one-loop contributions
are shown as a function of the cosine scattering angle
for the c.m. energies (\ref{set_energy})
in the $\alpha(0)$ EW scheme. The (1) stands for the fully polarized case,
while (2) -- for partially polarized case with
$(P_{e^+},P_{e^-}) = (+0.3,-0.8)$, and (3) -- for $(P_{e^+},P_{e^-}) = (0,-0.8)$.
The corresponding shift of the asymmetry
$$\Delta \alr = \alr ({\rm NLO\ EW}) - \alr ({\rm LO})$$
is shown in the lower panel.

We observe a symmetric
behavior of $\alr$ with respect to $\cos \vartheta_Z = 0$,
the significant dependence on
energy, the flatter behaviour with decreasing energy,
the large sensitivity to electroweak interaction effects and
to the degree of the initial beam polarization.

\subsection{Narrow-width approximation for the decay channel}

In this approach we create a narrow-width cascade
using Born and one-loop $e^+e^- \to Z Z$ and $Z \to \mu^+ \mu^-$ formulae, i.e.,
\begin{eqnarray}
\sigma^{\rm LO, NLO}_{e^+e^- \to 2\mu^+ 2\mu^-} = \sigma^{\rm LO, NLO}_{e^+e^- \to Z Z}\  
{\rm Br}^{\rm LO, NLO}_{Z \to \mu^+ \mu^-}\ {\rm Br}^{\rm LO, NLO}_{Z \to \mu^+ \mu^-},
\end{eqnarray}
where ${\rm Br}$ is a partial branching factor
\begin{eqnarray}
{\rm Br}^{\rm LO, NLO}_{Z \to \mu^+ \mu^-} = 
\frac{\Gamma^{\rm LO, NLO}_{Z \to \mu^+ \mu^-}}{\Gamma_Z}.
\end{eqnarray}

At one-loop, it is more consistent to use instead its “linearized” version
\begin{eqnarray}
\sigma^{\rm NLO}_{e^+e^- \to 2\mu^+ 2\mu^-} = \sigma^{\rm LO}_{e^+e^- \to Z Z}\  
{\rm Br}^{\rm LO}_{Z \to \mu^+ \mu^-} (1 + \delta_{e^+e^- \to Z Z} + 2\delta_{Z \to \mu^+ \mu^-}).
\end{eqnarray}

The partial width for $Z \to \mu^+ \mu^-$ decay at the LO and NLO levels are: 
$\Gamma^{\rm Born}=80.9363$ MeV, $\Gamma^{\rm NLO}=83.3286$ MeV, 
($\delta_{Z \to \mu^+ \mu^-} = 2.956 \%$)
and the corresponding branching factors are:
${\rm Br}^{\rm LO}_{Z \to \mu^+ \mu^-}$ = 0.032437,
${\rm Br}^{\rm NLO}_{Z \to \mu^+ \mu^-}$ = 0.033396.

Since the narrow-width approximation is valid at the 
c.m. energy of the reaction threshold $\sqrt{s} = 2M_Z \approx  182$ GeV,
the only lowest energy $\sqrt{s} = 250$ GeV of the accelerators is given
in Table~\ref{table:250nw}.

\begin{center}
\begin{table}[ht]
\caption{\label{table:250nw}
Integrated Born and one-loop cross sections
for unpolarized initial beams 
at the c.m. energy $\sqrt{s} = 250$ GeV for $e^+e^- \to ZZ 
\stackrel{\text{n.w.}}{\to} 2\mu^+ 2\mu^-$ channel
}
\begin{tabular}{lccccccc}
\hline\hline
$P_{e^+}, P_{e^-}$           & $0,0$         &$-1,+1$    & $+1,-1$  & $+0.3,-0.8$ & $-0.3,+0.8$ & $0,-0.8$  & $0,+0.8$ \\
\\
\hline 
$\sigma^{\rm Born}$, fb      & 1.0730(1)  & 1.2700(1) & 3.0220(1) & 1.8123(1) & 0.8487(1) & 1.4235(1) & 0.7226(1)\\
$\sigma^{\rm NLO}$, fb       & 1.1250(1)  & 1.6414(1) & 2.8579(1) & 1.7296(1) & 1.060(1) & 1.3685(1) & 0.8818(1)\\
$\delta^{\rm NLO}$, \%       & $4.85(1)$ & 29.25(1)  &$-5.43(1)$&$-4.57(1)$ & 24.95(1)   & $-3.86(1)$ & 22.04(1)\\
$\delta^{\rm NLO}_{\rm lin}$, \% & $4.83(1)$ & 27.84(1)  &$-4.87(1)$&$-4.06(1)$ & 23.79(1)& $-3.39(1)$ & 21.05(1)\\
\hline\hline
\end{tabular}
\end{table}
\end{center}

To obtain the differential distributions as well as final lepton polarization,
full off-shell/double-pole approximation calculations should be carried out. 
 
\section{Conclusion}\label{sect_Concl}

In this paper we have described the evaluation of polarization effects
for the cross sections of the $Z$ pair productions
at the one-loop level at high energies.

Comparisons of the results at the tree level for the Born and hard photon bremsstrahlung with 
{\calchep}~\cite{Belyaev:2012qa} and {\whizard}~\cite{Kilian:2007gr,Kilian:2018onl} are given, and a very good agreement is found.
Our numerical results for one-loop contributions fully confirm the results
in~\cite{Denner:1988tv}
and do not confirm the results of~\cite{Demirci:2022lmr}.

The angular and energy dependence with the effect of polarization of the initial
states was carefully analyzed for certain helicity states. 
The polarization effects were found to be significant.
The increase of the cross section for mostly negative electron
polarization compared to the unpolarized one was found.
The radiative corrections themselves were rather sensitive 
to degrees of polarization of the initial beams and depended quite strongly on energy.

In addition, calculations in the $\alpha(0)$ and $G_\mu$ EW schemes were considered. 
The results for relative corrections in the $G_\mu$ EW scheme are approximately 5-6\%
less than in the $\alpha(0)$ one.
The difference between complete one-loop cross sections in the considered EW schemes is about 1\% or less. 
This could be regarded as a theoretical uncertainty.

\section{Funding}
\label{sec:funding}

The research was supported by the Russian Science Foundation, project No. 22-12-00021.
\label{sec:acknowledgements}

\bibliographystyle{utphys_spires}
\bibliography{eeZZ}

\end{document}